\documentclass[a4paper,11pt]{article}
\usepackage{pos}
\usepackage{bm}
\usepackage{amsmath}
\usepackage{hyperref}
\usepackage{comment}
\usepackage{enumitem}

\newcommand{\CC}{\mathbb{C}}
\newcommand{\RR}{\mathbb{R}}

\newcommand{\NN}{\mathbb{N}}

\newcommand{\la}{\langle}
\newcommand{\ra}{\rangle}

\newcommand{\1}{{\bf 1}}
\newcommand{\Hi}{{\cal H}}

\newcommand{\A}{{\cal A}}
\newcommand{\B}{{\cal B}}
\newcommand{\C}{{\cal C}}
\newcommand{\E}{{\cal E}}
\newcommand{\F}{{\cal F}}
\newcommand{\I}{{\cal I}}
\newcommand{\J}{{\cal J}}
\newcommand{\T}{{\cal T}}

\newcommand{\Ps}{{\cal P}}
\newcommand{\Pt}{{\cal P}^t}
\newcommand{\Pe}{{\sf P}}

\newcommand{\Ss}{{\cal S}}

\newcommand{\bx}{{\bm x}}
\newcommand{\bD}{{\bf D}}
\newcommand{\bL}{{\bm L}}
\newcommand{\btheta}{{\bm\theta}}
\newcommand{\blambda}{{\bm \lambda}}
\newcommand{\bLambda}{{\bm \Lambda}}
\newcommand{\bpi}{{\bm\pi}}
\newcommand{\bphi}{{\bm\phi}}
\newcommand{\bpsi}{{\bm\psi}}

\newcommand{\be}{\begin{equation}}
\newcommand{\ee}{\end{equation}}
\newcommand{\bea}{\begin{eqnarray}}
\newcommand{\eea}{\end{eqnarray}}
\newcommand{\ba}{\begin{array}}
\newcommand{\ea}{\end{array}}

\title{General $O(D)$-equivariant fuzzy hyperspheres via confining potentials and energy cutoffs}
\ShortTitle{General $O(D)$-equivariant fuzzy hyperspheres}

\author*[a,b]{Gaetano Fiore }

\affiliation[a]{Dip. di Matematica e applicazioni, Universit\`a di Napoli ``Federico II'',\\
Complesso Universitario  M. S. Angelo, Via Cintia, 80126 Napoli, Italy}

\affiliation[b]{INFN, Sezione di Napoli,\\
Complesso Universitario  M. S. Angelo, Via Cintia, 80126 Napoli, Italy}

\emailAdd{gaetano.fiore@na.infn.it}

\abstract{We summarize our recent construction \cite{FioJPA23,FioPis18,Pis20} of new fuzzy hyperspheres $S^d_{\Lambda}$ of arbitrary dimension $d\in\NN$ covariant under the {\it full} orthogonal group $O(D)$, $D=d\!+\!1$. We impose a suitable energy cutoff on a quantum particle in $\RR^D$ subject to a confining potential well $V(r)$  
with a very sharp minimum on the sphere of radius $r=1$; the cutoff and the depth of the well diverge with $\Lambda\in\NN$.  Consequently, the commutators  of the Cartesian coordinates $\overline{x}^i$ are  proportional to the angular momentum components $L_{ij}$, as in Snyder's noncommutative spaces. The $\overline{x}^i$ generate the whole algebra of observables  $\A_{\Lambda}$ and thus the whole Hilbert space $\Hi_{\Lambda}$ when applied to any state.
$\mathcal{H}_{\Lambda}$ carries a reducible representation of $O(D)$  isomorphic  to  the space  
of harmonic homogeneous polynomials of degree $\Lambda$ in the Cartesian coordinates of (commutative) $\RR^{D+1}$; the latter carries an irreducible representation $\bpi_\Lambda$  of  $O(D\!+\!1)\supset O(D)$.
Moreover, $\A_{\Lambda}$  is isomorphic to $\bpi_\Lambda\left(Uso(D\!+\!1)\right)$. 
We identify the subspace $\C_\Lambda\subset\A_{\Lambda}$ spanned by fuzzy spherical harmonics. We interpret $\{\Hi_\Lambda\}_{\Lambda\in\NN}$, $\{\C_\Lambda\}_{\Lambda\in\NN}$ as fuzzy deformations of
the space  $\Hi_s\equiv {\cal L}^2(S^d)$ of square integrable functions 
and the space  $C(S^d)$ of continuous functions  on $S^d$ respectively,  
$\{\A_\Lambda\}_{\Lambda\in\NN}$ as fuzzy deformation of 
 the associated algebra $\A_s$ of observables, because they resp. go to  $\Hi_s,C(S^d),\A_s$
 as $\Lambda$ diverges (with fixed $\hbar$).
With suitable $\hbar=\hbar(\Lambda)\stackrel{\Lambda\to\infty}{\longrightarrow} 0$,
in the same limit $\A_\Lambda$ goes to the (algebra of functions on the) Poisson manifold $T^*S^d$; 
 more formally,  $\{\A_\Lambda\}_{\Lambda\in\NN}$ yields a fuzzy quantization of a coadjoint orbit of $O(D\!+\!1)$ that goes to the classical phase space $T^*S^d$. 
These models might be useful in quantum field theory,  quantum gravity or condensed matter physics.}

\FullConference{%
  Corfu Summer Institute 2022 "School and Workshops on Elementary Particle Physics and Gravity",\\
  28 August - 1 October, 2022, \\ 
``Workshop on Noncommutative and generalized geometry in string theory, gauge theory and related physical models”, 18-25 September, 2022,\\
  Corfu, Greece}

\begin{document}
\maketitle

\section{Introduction and preliminaries}
\label{intro}

In the past decades noncommutative space(time) algebras have been introduced and studied 
as fundamental or effective arenas for \  regularizing ultraviolet (UV) divergences in quantum field theory (QFT) (see e.g. \cite{Snyder}), 
\ reconciling Quantum Mechanics and General Relativity in a satisfactory Quantum Gravity (QG) theory   (see e.g. \cite{DopFreRob95}), 
\ unifying fundamental interactions  (see e.g. \cite{ConLot90,ChaCon10}).
Noncommutative Geometry  (NCG) \cite{Connes,Lan97,Madore99,GraFigVar00},  i.e. differential geometry  on noncommutative spaces, has become a sophisticated machinery. 
In particular, fuzzy (noncommutative) spaces  have raised a big interest as a non-perturbative technique in  QFT  based on a finite discretization alternative to the lattice ones. A fuzzy space is a sequence $\{\mathcal{A}\}_{n\in\NN}$ of {\it finite-dimensional} algebras such that
 $\mathcal{A}_n\overset{n\rightarrow\infty}\longrightarrow\mathcal{A}\!\equiv$algebra 
of regular functions on an ordinary manifold, with \ dim$(\mathcal{A}_n)\overset{n\rightarrow\infty}\longrightarrow\infty$. Contrary to lattices,
$\A_n$ can carry representations of Lie, beside discrete, groups. Fuzzy spaces can be used also  to discretize internal (e.g. gauge) degrees of freedom (see e.g. \cite{AscMadManSteZou}),
or  as a new tool in string and $D$-brane theories (see e.g. \cite{AleRecSch99,HikNozSug01}). In the  seminal Madore-Hoppe  Fuzzy Sphere (FS) of dimension $d=2$ \cite{Mad92,HopdeWNic}  
${\cal A}_n\simeq M_n(\CC)$. ${\cal A}_n$  is generated by coordinates $x^i$ ($i=1,2,3$) fulfilling
\be
\ba{l}
[x^i,x^j]=\frac {2i}{\sqrt{n^2\!-\!1}}\varepsilon^{ijk}x^k, \qquad
r^2\equiv x^ix^i=1,\qquad\quad n\in\NN\setminus \{1\};  
\ea                   \label{FS}
\ee
%(sum over repeated indices);
these are related via $x^i=2L_i/{\sqrt{n^2\!-\!1}}$ to the standard basis  $\{L_i\}_{i=1}^3$ of $so(3)$ in the unitary irreducible representation (irrep)  $(\pi^l,V^l)$ of dimension   $n=2l\!+\!1$ [i.e. $V^l$ is the eigenspace of the Casimir $\bL^2=L_iL_i$ with eigenvalue $l(l+1)$]. 
%Ref. \cite{GroMad92,GroKliPre96'} first proposed a QFT based on it.
Fuzzy spheres $S^d$ of dimension $d=4$ and any $d\ge 3$ were introduced resp. in \cite{GroKliPre96}, \cite{Ramgoolam}; other versions of $d=3,4$ or $d\ge 3$ 
 in \cite{DolOCon03,DolOConPre03,Ste16,Ste17}.
Unfortunately, while for the $d=2$ FS \cite{Mad92,HopdeWNic}  ${\cal A}_n$ admits  a basis of spherical harmonics,
% (polynomials in the $x^i$ transforming under $SO(3)$ as spherical harmonics on $S^2$), 
for the $d>2$ fuzzy  $S^d$ a product of spherical harmonics is not a combination thereof, but an element in a larger algebra $\A_n$.

The Hilbert space of a (zero-spin) quantum particle on configuration space $S^d$ and the space of continuous functions on $S^d$ carry a (same)  {\it reducible} representation of 
$O(D)$,  $D\equiv d\!+\!1$; they decompose into carrier spaces of irreducible representations (irreps) as follows
\bea
{\cal L}^2(S^d)\simeq\bigoplus\nolimits_{l=0}^\infty V_D^l\simeq C(S^d),
\label{directsum}
\eea
where  $V_D^l$ is an eigenspace of the quadratic Casimir $\bL^2$  with eigenvalue  
\bea
E_l\equiv l(l\!+\!D\!-\!2)           \label{El}
\eea
($V_3^l\equiv V^l$); $C(S^d)$ acts an algebra of bounded operators on ${\cal L}^2(S^d)$.
  On the contrary, each of the mentioned fuzzy hyperspheres is  based on a sequence parametrized by $n$ either of irreps of $Spin(D)$ (so that $r^2\propto\bL^2$  is 1)  \cite{Mad92,HopdeWNic,GroKliPre96,Ramgoolam,DolOCon03,DolOConPre03},
or %on a sequence of reducible representations that are the 
of direct sums of small bunches of  such irreps \cite{Ste16,Ste17}. In either case, even excluding the $n$'s for which the associated representation of $O(D)$ is only {\it projective},
% [$Spin(D)$ is the universal cover of $SO(D)$, not of $O(D)$], 
the carrier space does not go to (\ref{directsum})
as $n\to\infty$; hence, interpreting these fuzzy spheres
as fuzzy configuration spaces $S^d$ (and the $x^i$ as spatial coordinates) becomes questionable.
Moreover, relations  (\ref{FS}) for the Madore-Hoppe FS are equivariant under $SO(3)$, but not under the whole $O(3)$, e.g. not under parity $x^i\mapsto -x^i$.
%, while the ordinary sphere $S^2$ is (on the contrary, all the other mentioned fuzzy spheres  are $O(D)$-equivariant, because $[x^i,x^j]\propto L_{ij}$).
These difficulties are overcome by our recent fully $O(D)$-equivariant  fuzzy quantizations \cite{FioJPA23,Pis20} $S^d_{\Lambda}$ of spheres  $S^d$ of arbitrary dimension $d=D\!-\!1\in\NN$ (thought as configuration spaces) and of $T^*S^d$
(thought as phase spaces), which we summarize here (the cases $d=1,2$ had been treated in \cite{FioPis18,FioPis18POS});
%in a sense that will be fully clarified at the end of section \ref{discuss} overcomes these difficulties; 
in particular, we recover (\ref{directsum}) as $\Lambda\to\infty$. 

Our  fuzzy quantization  uses: %two ingredients: 
1. the {\it projection} of a quantum theory  $\T$ on $\RR^D$ below an {\it energy cutoff}; 2. a {\it dimensional reduction} induced by a  {\it confining potential} on $S^d\subset\RR^D$. One can apply it to quantize also other submanifolds $M\subset\RR^D$.
Given a generic quantum theory $\T$ with Hilbert space  $\Hi$,  algebra of observables on $\Hi$ (or with a domain dense in $\Hi$) $\A\equiv\mbox{Lin}(\Hi)$, Hamiltonian   $H\in\A$, for any subspace $\overline{\Hi}\subset\Hi$  preserved by  $H$ let $\overline{P}:\Hi\mapsto\overline{\Hi}$ \ be  the associated projector and
$$
\overline{\A}\equiv \mbox{Lin}\big(\overline{\Hi}\big)
=\{\overline{A}\equiv \overline{P}A\overline{P}\:\: |\: A\in\A\}.
$$
By construction   $\overline{H}=\overline{P}H=H\overline{P}$.   
The projected Hilbert space $\overline{\Hi}$, algebra of observables $\overline{\A}$ and Hamiltonian  $\overline{H}$  provide a new quantum theory $\overline{\T}$ \cite{FioPis20PoS};  
we will ascribe the observable $\overline{A}$  the same physical meaning of $A$ in $\T$.
If $\overline{\Hi}$, $H$ are invariant under some  group $G$, then 
\ $\overline{P},\overline{\A},\overline{H},\overline{\T}$ \ will be as well.
The relations among the generators of $\overline{\A}$ differ from those among the generators of $\A$. In particular,  if  $\T$ is based on commuting coordinates $x^i$
(commutative space) this will be in general no longer  true for $\overline{\T}$: 
\ $[\overline{x}^i,\overline{x}^j]\neq 0$, \ and we have generated a quantum theory on a NC space. In particular, if
$\overline{\Hi}\subset\Hi$ is characterized by energies  $E\le\overline{E}$ below
a certain cutoff $\overline{E}$, then
$\overline{\T}$ is a low-energy approximation  of $\T$ preserved by the dynamical evolution ruled by $H$. $\overline{\T}$ may be used as an effective theory for $E\le\overline{E}$, or may even help to figure out a new theory $\T'$ valid for all $E$ if at $E>\overline{E}$  physics is not accounted for by $\T$. If $\overline{\T}$ describes an ordinary (for simplicity, zero-spin) quantum particle in the Euclidean (configuration) space $\RR^D$, then $\Hi={\cal L}^2(\RR^d)$. If $H=T\!+\!V$, with  kinetic energy $T$ and a confining potential $V(x)$, then the classical region $\B_{\scriptscriptstyle\overline{E}}$ in  phase space fulfilling $H(x,p)\le \overline{E}$ and the one $v_{\overline{E}}\subset \RR^D$ in configuration space
 fulfilling  \ $V\le \overline{E}$ \ are bounded  at least for sufficiently small  $\overline{E}$, and  the dimension $\mbox{dim}(\overline{\Hi})\approx\mbox{Vol}(\B_{\scriptscriptstyle\overline{E}})/h^D$ of $\overline{\Hi}$ is finite. 
In the sequel we rescale $x,p,H,V$ so that they are dimensionless and, denoting by
$\Delta$ the Laplacian in  $\RR^D$, 
\be
H=-\Delta + V.                                      \label{Ham}
\ee
We choose a sequence of pairs $(V,\overline{E})$ satisfying the following requirements. $V=V(r)$ has a very sharp minimum, parametrized by a very large $k\equiv V''(1)/4$,  on the sphere $S^d\subset\RR^D$ of radius $r=1$; we fix $V_0\equiv V(1)$ so that the ground state $\bpsi_0$ has zero energy, $E_0=0$.  We  choose  $\overline{E}$
fulfilling first of all the condition $V(r)\simeq V_0\!+\!2k (r\!-\!1)^2%\label{cond1}
$ in $v_{\overline{E}}$, so that we can approximate $v_{\overline{E}}$  by  the spherical shell
$|r\!-\!1|\le \sqrt{\frac{\overline{E}\!-\!V_0}{2k}}$ and the potential by a harmonic one.
If $\overline{E}\!-\!V_0$ and $k$ diverge, while their ratio goes to zero, then in this limit $v_{\overline{E}}\to S^d$, dim$(\overline{\Hi})\to\infty$, and we recover quantum mechanics on  $S^d$. 

Let $x\!\equiv\! (x^1\!,\!...x^D)$ be Cartesian coordinates of $\RR^D$,  $r^2\!=\!x^ix^i$, $\partial_i\!\equiv\! \partial/\partial x^i$;  $\Delta\!=\! \partial_i\partial_i$  decomposes as %follows
\be
\Delta=\partial_r^2 \:+\: (D\!-\!1)\, r^{-1}\partial_r \:-\: r^{-2} \bL^2,  \label{LaplacianD}
\ee
where $\partial_r\!\equiv\! \partial/\partial r$ and $\bL^2\equiv L_{ij}L_{ij}/2$ is the square angular momentum 
(in normalized units), i.e. the quadratic Casimir of $Uso(D)$ and the Laplacian on the sphere $S^d$,  the angular momentum components $L_{ij}\!\equiv\! i(x^j\partial_i-x^i\partial_j)$ are  
%$\bL^2$ can be also expressed in terms of angles $\theta_a$ and derivatives $\partial/\partial \theta^a$. 
vector fields tangent to all spheres $r =$ const  satisfying 
\bea
[L_{ij},L_{hk}]=i\left(L_{jk}\delta_{hi}+L_{ih}\delta_{kj}-L_{jh}\delta_{ki}-L_{ik}\delta_{hj}\right), \qquad [L_{ij},S]= 0,    \label{Lcr} \\[2pt]
{} [iL_{ij},v^h]= v^i\delta^h_j-v^j\delta^h_i, \qquad\qquad
\varepsilon^{i_1i_2i_3....i_D}x^{i_1}L_{i_2i_3}=0,           \label{Lijrel}
\eea 
where $S$ is any scalar and $v^h$ are the components  of any vector depending on $x^h,\partial_h$, in particular $v^h=x^h,\partial_h$. 
%$L_{ij}$ generates rotations in the $x^ix^j$ plane.
The Ansatz \ $\bpsi=f(r)Y_l(\btheta)$, \ with $f(r)= r^{-d/2} g(r)$ and $Y_l\!\in\! V_D^l$ an $E_l$-eigenfunction of $\bL^2$,  transforms the Schr\"odinger PDE \ $H\bpsi=E\bpsi$ \ into the Fuchsian ODE in the unknown  $g(r)$
\be
\ba{l}
-g''(r)+\left[V(r)+\frac{D^2-4D+3+4l(l+D-2)}{4}\,r^{-2}\right]g(r)=Eg(r)
\ea
\label{eqpolarD2}
\ee
(by similar product Ans\"atze one can reduce numerous different PDEs to ODEs, see e.g. \cite{DeAFio13}).
Requiring $\lim_{r\rightarrow 0^+}r^2V(r)>0$, %$r^2V(r)\overset{r\rightarrow 0^+}\longrightarrow T\in\mathbb{R}^+$, 
$f(0)=0$,
we make $H$ self-adjoint. As $V(r)$ is very large outside  $v_{\overline{E}}$, there $g,f,\bpsi$ are negligibly small,  and  the lowest eigenvalues $E$ are at leading order those of the $1$-dimensional harmonic oscillator approximation \cite{Pis20} of (\ref{eqpolarD2})
\be
-g''(r)+g(r)k_{l}\left(r-\widetilde{r}_{l}\right)^2=\widetilde{E}_{l}g(r),
\label{eqpolarD3}
\ee
obtained neglecting terms $O\big((r\!-\!1)^3\big)$ in the Taylor expansions of
$1/r^2,V(r)$ about $r\!=\!1$. Here 
$$\label{definizioni1}
\ba{l}
\widetilde{r}_{l}\equiv 1+\frac{b(l,D)}{3b(l,D)+2k },\quad \widetilde{E}_{l}\equiv E-V_0\frac{2b(l,D)\left[k +b(l,D)\right]}{3b(l,D)+2k }, \\[6pt]
k_{l}\equiv 2k+3b(l,D) ,\quad 
b(l,D)\equiv \frac{D^2-4D+3+4l(l+D-2)}{4}.
\ea  
$$
The square-integrable solutions of (\ref{eqpolarD3})  $g_{n,l}(r)$  lead to
\be\label{valuef}
f_{n,l}(r)=M_{n,l}\hspace{0.15cm} r^{-d/2}\: e^{-\sqrt{k_{l}}\left(r-\widetilde{r}_{l}\right)^2/2}\cdot H_n\left((r-\widetilde{r}_{l})\sqrt[4]{k_{l}}\right)\quad\mbox{ with }n\in\mathbb{N}_0;
\ee
here $M_{n,l}$ are   normalization constants and $H_n$ are the Hermite polynomials.
The corresponding `eigenvalues' in (\ref{eqpolarD3})   
$\widetilde{E}_{n,l}=(2n+1)\sqrt{k_{l}}$   lead to energies
$E_{n,l}=(2n+1)\sqrt{k_{l}}+V_0+\frac{2b(l,D)[k +b(l,D)]}{3b(l,D)+2k }$. 
As said, we fix $V_0$ requiring that the lowest one $E_{0,0}$ be zero; this implies
$V_0=-\sqrt{2k }-b(0,D)-\frac{3b(0,D)}{2\sqrt{2k }}+O\big(k ^{-1/2}\big)$,
and the expansions of  $E_{n,l}$ and $\widetilde{r}_{l}$ at leading order in $k $ become
\be 
E_{n,l}={\color{blue}{l(l+D-2)}}+ {\color{red}{2n\sqrt{2k}}}+O\big(k ^{-2}\big),\qquad
\widetilde{r}_{l}={\color{blue}{1}}+ b(l,\!D)/2k +O\big(k ^{-2}\big).
\ee 
$E_{0,l}$ coincide at lowest order with  the desired eigenvalues $E_l$ (coloured blue)  of $\bL^2$, while  if $n>0$
$E_{n,l}$ diverge as $k \to\infty$ (due to the red part);  to exclude all states with $n>0$ (i.e.,  to `freeze' radial oscillations, so that all corresponding classical trajectories are circles; this can be considered as a {\it quantum} version of the  {\it constraint} $r=1$) we  impose the energy cutoff
\bea
E_{n,l}\le \overline{E}(\Lambda)\equiv \Lambda(\Lambda\!+\!D\!-\!2) <2\sqrt{2k},\qquad\Lambda\in\NN. \label{consistencyD}
\eea
The right inequality is satisfied prescribing a suitable dependence $k \left(\Lambda\right)$, e.g. $k \left(\Lambda\right)\equiv\left[\Lambda(\Lambda\!+\!D\!-\!2) \right]^2$; the left one  is satisfied  setting $n=0$ and $l\le \Lambda$.  We rename $\overline{H},\overline{\Hi},\overline{P},\overline{\A},\overline{\T}$ as  
$H_{\Lambda},\Hi_{\Lambda},P_{\Lambda},\A_{\Lambda},\T_{\Lambda}$. $\T_{\Lambda}$ is $O(D)$-equivariant. 
We end up with eigenfunctions  and eigenvalues (at leading order in $1/\Lambda$)
\be
\bpsi_l(r,\btheta)= f_l(r)\,Y_l(\btheta), \qquad H_{\Lambda}\bpsi_l=E_{l}\,\bpsi_l,
\qquad\quad
l=0,1,...,\Lambda,\label{statopsi}
\ee
abbreviating   $ f_l\equiv f_{0,l}$. 
Hence $\Hi_{\Lambda}$ decomposes into irreps of $O(D)$ (and eigenspaces of $\bL^2,H_{\Lambda}$) as 
\be
\Hi_{\Lambda}=\bigoplus\nolimits_{l=0}^{\Lambda}\Hi_{\Lambda}^l, \qquad  \Hi_{\Lambda}^l\equiv f_l(r)\, V_D^l.    \label{decoHiLambda}
\ee
%In the limit 
As $\Lambda \to\infty$ the spectrum $\{E_l\}_{l=0}^{\Lambda}$ of $H_{\Lambda}$ goes to the whole spectrum $\{E_l\}_{l\in\mathbb{N}_0}$ of   $\bm{L}^2$, and we recover (\ref{directsum}). 
We can express the projectors $P^l_{\Lambda}:\Hi_{\Lambda}\to\Hi_{\Lambda}^l$  as the following 
polynomials in $\overline{\bL}^2$:
\be
\ba{l}
P^l_{\Lambda}=\prod_{n=0, n\neq l}^{\Lambda} \frac{\overline{\bL}^2-E_n}
{E_l-E_n}. 
\ea   \label{Proj-l}
\ee
\begin{comment}
Let $\la T,T'\ra\equiv \int_{S^d}\!  d\alpha\: T^*\,T'$ be the scalar product of $\mathcal{L}^2(S^d)$; here $d\alpha$ is the $O(D)$-invariant measure on $S^d$\footnote{$d\alpha=\varepsilon^{i_1...i_D}x^{i_1}dx^{i_2}... dx^{i_D}$, or equivalently $d\alpha=\sin^{d-1}{(\theta_{d})} \sin^{d-2}{(\theta_{d-1})} \cdots \sin{(\theta_2)}... d\theta_1 d\theta_2\cdots d\theta_{d}$.
}.
If  $\bphi,\bphi'\in\Hi\equiv {\cal L}^2(\mathbb{R}^D)$ 
can be factorized  into radial parts $f(r),f'(r)$ and  angular parts $T,T'\in\mathcal{L}^2(S^d)$, i.e. $\bphi=f\,T$, $\bphi'=f'\,T'$, then so can be their scalar product:
\bea
\left(\bphi,\bphi'\right)\equiv \int_{\mathbb{R}^D}d^Dx\,\bphi^*(x)\bphi'(x)
=\la T,T'\ra\: \int^\infty_0\!\!\! dr\,r^d\,f^*(r) \, f'(r).
\label{scalprodRD}
\eea
\end{comment}
The space $V_D^l$ consists of harmonic homogeneous polynomials of degree $l$ in the $x^i$ restricted to the sphere $S^d$. In  section \ref{xt-polynomials} we show: i) how
to explicitly determine $V_D^l$, as well as the action of $L_ {hk}$ and $t^h\equiv x^h/r$ on $V_D^l$, applying the trace-free completely symmetric projector $\Ps^l$ of $\left(\RR^D\right)^{\otimes^l}$ to the homogeneous polynomials of degree $l$ in   $x^i$;
%; this allows to avoid the rather complicated  actions of the $L_{hk}$ on spherical harmonics (which also span $V_D^l$) used in \cite{Pis20}.
ii) that not only $\Hi_{\Lambda}$, but also $V_{D+1}^{\Lambda}$ decomposes into irreps of  $O(D)$ as follows \ 
$%\Hi_{\Lambda}\simeq 
V_{D+1}^{\Lambda}\simeq\bigoplus\nolimits_{l=0}^{\Lambda} V_D^l. %\label{Hi_Lambda-deco}\ee
$
In section \ref{xLRel} we write down the 
 relations fulfilled by   $\overline x^i,\overline L_{hk}$ and point out that:  the $*$-algebra $\A_{\Lambda}$ generated by the latter is also generated by the $\overline x^i$ alone; ii) the unitary irrep of  $\A_{\Lambda}$ on $\Hi_{\Lambda}$ is 
isomorphic to the irrep $\bpi_\Lambda$ of $Uso(D+1)$ on $V_{D+1}^{\Lambda}$.
In section \ref{Dconverge} we show in which sense  $\Hi_{\Lambda},\A_{\Lambda}$
go to $\Hi,\A$ as $\Lambda\to\infty$, in particular how one can recover  the multiplication operator $f\cdot\in C(S^d)\subset\A$ of wavefunctions in ${\cal L}^2(S^d)$ by a continuous function $f$ as the strong limit of a suitable sequence $f_\Lambda\in\A_\Lambda$.
In section \ref{discuss} we discuss our results and possible developments in comparison with the literature; in particular, we point out 
that with a suitable $\hbar(\Lambda)$ our pair  $(\Hi_{\Lambda},\A_{\Lambda})$ can be seen as a fuzzy quantization
of a coadjoint orbit of $O(D)$ that can be identified with the cotangent space $T^*S^d$,
the classical phase space over the $d$-dimensional sphere.

\section{Representations of $O(D)$  via polynomials in $x^i$, $t^i\equiv x^i/r$}
\label{xt-polynomials}

Let \ $\CC[x^1,...,x^D]=\bigoplus_{l=0}^{\infty}W_D^l$   be the decomposition of the space of complex polynomial functions on $\RR^D$ into 
subspaces $W_D^l$ of homogeneous ones of degree $l$.
\begin{comment}
The monomials  $x^{i_1}x^{i_2}...x^{i_l}$  can be reordered in the form $(x^1)^{l_1}...(x^D)^{l_D}$ and make up a basis  of $W_D^l$:
$$
\B_{W_D^l}\equiv \left\{(x^1)^{l_1}...(x^D)^{l_D}\:| \: (l_1,...,l_D)\in\NN_0, \: \sum_{i=1}^Dl_i=l\right\}, \qquad\mbox{dim}(W_D^l)={D\!+\!l\!-\!1\choose l};
$$
 the dimension of $W_D^l$  is computed as the number of elements of $\B_{W_D^l}$. 
\end{comment}
If $l\ge 2$ then $W_D^l$ carries a reducible representation of  $O(D)$, as well as $Uso(D)$, because by (\ref{Lcr}b) the subspace $r^2W_D^{l-2}\subset W_D^l$ carries a smaller one.  \begin{comment} 
The `trace-free' component $\check V_D^{l}$ of  $W_D^l$, consisting of harmonic polynomials in $x^i$ (i.e.  annihilated by $\Delta$),  carries an irrep of  $O(D),Uso(D)$, and $W_D^l=r^2W_D^{l-2}\oplus \check V_D^l$. As a consequence,
\be
\mbox{dim}(\check V_D^l)=\mbox{dim}(W_D^l)-\mbox{dim}(W_D^{l-2})
=\frac{(l\!+\!D\!-\!3)...(l\!+\!1)}{(D\!-\!2)!}(D\!+\!2l\!-\!2).  \label{dimVDl} 
\ee
\end{comment}
The `trace-free' component $\check V_D^{l}$ in the decomposition $W_D^l=r^2W_D^{l-2}\oplus \check V_D^l$ carries the irrep $\bpi_D^{l}$ of $Uso(D)$ and $O(D)$  characterized by the highest eigenvalue of $\bL^2$ within $W_D^l$, namely $E_l$.
In fact,  for all $h,k\in\{1,...,D\}$
\ $X_{l,\pm}^{hk}\equiv(x^h\!\pm\!i x^k)^l\in W_D^l$ \ are eigenvectors of $\bL^2$ with  eigenvalue $E_l$, of $L_{hk}$ with eigenvalue $\pm l$, and of $\Delta$  with eigenvalue 0.
Hence \ $X_{l,+}^{hk}$, $X_{l,-}^{hk}$ \ can be used as the highest and lowest weight vectors of $(\bpi_D^{l},\check V_D^{l})$ \cite{FioJPA23}.
Since all the $L_{ij}$ commute with $\Delta$, $\check V_D^{l}$  can be characterized also as the subspace of $W_D^l$ that is annihilated by $\Delta$. 
A complete set in $\check V_D^{l}$ consists of trace-free homogeneous polynomials
$X_l^{i_1i_2...i_l}$, which we obtain below
applying the completely symmetric trace-free projector $\Ps^l$ to the monomials $x^{i_1}x^{i_2}...x^{i_l}$. 
We slightly enlarge $\CC[x^1,...x^D]$ by new scalar generators $r,r^{-1}$ fulfilling the relations \ $r^2=x^ix^i$, $rr^{-1}=1$.  Its elements 
\be
t^i\equiv r^{-1}x^i, \qquad\qquad T_{l,\pm}^{hk}\equiv (t^h\!\pm\!i t^k)^l=r^{-l}X_{l,\pm}^{hk}
\ee
% $t^i\equiv x^i/r$ \ 
fulfill the following relations: \ i) \ $t^it^i=1$, \ which
characterizes the coordinates of points of  $S^d$; hence $V_D^l\equiv r^{-l}\check V_D^{l}$ can be seen as the restriction of $\check V_D^{l}$ to $S^d$. ii) $T_{l,\pm}^{hk}\in V_D^l$ \ are eigenvectors of $\bL^2$ with  eigenvalue $E_l$ and of $L_{hk}$ with eigenvalue $\pm l$;  \ hence $T_{l,+}^{hk}$, $T_{l,-}^{hk}$ \ can be used as the highest and lowest weight vectors of $(\bpi_D^{l},V_D^{l})$.
We denote by $Pol_D$  the algebra of complex polynomials in the $t^i$, by $Pol_D^{\Lambda}$ the subspace
of polynomials of degree $\Lambda$, by $P^{\Lambda}:Pol_D\to Pol_D^{\Lambda}$ the corresponding projector.  $Pol_D$ endowed with the scalar product \ 
$\la T,T'\ra\equiv \int_{S^d}d\alpha\, T^*T'
$ \ is a pre-Hilbert space, whose completion is 
${\cal L}^2(S^d)$; here $d\alpha=\varepsilon^{i_1...i_D}x^{i_1}dx^{i_2}... dx^{i_D}$ is the $O(D)$-invariant measure on $S^d$. 
We  extend $P^{\Lambda}$ to all of ${\cal L}^2(S^d)$ by continuity
in the norm of the latter. Also $Pol_D^\Lambda, V_D^l$  are Hilbert subspaces of ${\cal L}^2(S^d)$.
$Pol_D^{\Lambda}=W_D^{\Lambda}r^{-\Lambda}\oplus W_D^{\Lambda-1}r^{1-\Lambda}$ carries a reducible representation of $O(D)$ [and $Uso(D)$]  that  splits 
into irreps as \ $Pol_D^{\Lambda}=\bigoplus_{l=0}^\Lambda  V_D^l$. 
One finds  
% $\mbox{dim}\left(Pol_D^{\Lambda}\right)=\mbox{dim}\left(V_{D+1}^\Lambda\right)$, \ and actually 
\ $\Hi_\Lambda\simeq Pol_D^{\Lambda}\simeq  V_{D+1}^\Lambda$ \
 as $Uso(D)$ representations. The first isomorphism
follows from (\ref{decoHiLambda}), the second from section \ref{Embeddings}.

\subsection{$O(D)$-irreps via trace-free  completely symmetric projectors}
\label{ComplSymProj}

Let $(\pi,\E)$ be the fundamental ($D$-dimensional irreducible unitary) representation of $Uso(D)$ and
$O(D)$; the carrier space $\E$ is isomorphic to $V_D^1$. As a vector space $\E\simeq \RR^D$;
the set of Cartesian coordinates $x\equiv (x^1,...x^D)\in\RR^D$ can be seen as the set of components
of an element of $\E$ with respect to (w.r.t.) an orthonormal basis. The permutator on \ 
$\E^{\otimes^2}\equiv \E \otimes \E$ \ is defined via $\Pe(u\otimes v)=v\otimes u$
and linearly extended. In all bases it is represented by the $D^2\times D^2$
matrix \  $\Pe^{hi}_{jk}=\delta^{h}_{k}\delta^{i}_{j}$. 
The symmetric and antisymmetric  projectors $\Ps^+,\Ps^-$ on $\E^{\otimes^2}$ are obtained as
\bea
\Ps^\pm =\frac 12\left(\1_{D^2}\pm \Pe\right).              \label{sym-antisym_projectors}
\eea
Here and below we denote by $\1_{D^l}%=\left( \1_{D^l}{}^{h_1...h_l}_{i_1...i_l}\right)
$ the identity operator on $\E^{\otimes^l}$; in all bases it is represented by the $D^l\times D^l$ matrix
$\1_{D^l}{}^{h_1...h_l}_{i_1...i_l}\equiv \delta^{h_1}_{i_1}...\delta^{h_l}_{i_l}$.
\  $\Ps^-\E^{\otimes^2}$ carries an irrep under $O(D)$, while $\Ps^+\E^{\otimes^2}$  is the direct sum of two irreps:
the 1-dimensional {\it trace} and the $\frac12(D\!-\!1)(D\!+\!2)$-dimensional {\it trace-free symmetric} ones. The associated projectors $\Pt, \Ps^s$ from  $\E^{\otimes^2}$ are   given by
\be 
\Pt{}_{kl}^{ij} = \frac 1{D} \delta^{ij}\delta_{kl},    \qquad \Ps^s=\Ps^+-\Pt=\frac 12\left(\1_{D^2}+ \Pe\right)-\Pt \:;    \label{Pts} 
\ee 
here and below we adopt an orthonormal basis of $\E$ for the matrix representation of  $\Pt$. Hence $\Pt{}_{kl}^{ij}x^ix^j=\delta^{ij}r^2/D$.
These projectors satisfy the equations $\Ps^{\alpha}\Ps^{\beta} = \Ps^\alpha \delta^{\alpha\beta}, \quad \sum_\alpha\Ps^{\alpha}= \1_{D^2}$,
where $\alpha,\beta = -,s,t$. 
%This implies in particular \ $\Ps\Ps^{'}=0$, \ where 
%we have introduced the new projector \ $\Ps^{'}\equiv \Ps^{-}+\Pt$. \
$\Pe,\Pt$ are symmetric matrices, i.e. invariant under transposition
${}^T$, and therefore also the other projectors are, 
$\Pe^T=\Pe$, $\Ps^{\alpha}{}^T={\cal P}^{\alpha}$.
 In the sequel we abbreviate $\Ps\equiv\Ps^s$. 
Given a (linear) operator $M$  on  $\E^{\otimes^n}$,
for all integers $l,h$ with $l>n$,  and $1\le h\le l\!+\!1\!-\!n$ we denote by 
$M_{h(h\!+\!1)...(h\!+\!n\!-\!1)}$ the operator  on  $\E^{\otimes^l}$ acting
as the identity on the first $h\!-\!1$ and the last $l\!+\!1\!-\!n\!-\!h$ tensor factors,
and as $M$ in the remaining central ones. For instance, if $M=\Pe$ and $l=3$ 
we have $\Pe_{12} = \Pe \otimes \1_D$,  
$\Pe_{23} =  \1_D\otimes \Pe $.
All the projectors 
$A=\Ps^+,\Ps^-,\Ps,\Pt$ fulfill the relations  
\be 
A_{12}\,\Pe_{23}\,\Pe_{12}  
=\Pe_{23}\,\Pe_{12}\,A_{23},   \label{braid1} 
\ee 
\bea
D\: \Pt_{23}\Pt_{12}=\Pe_{12}\Pe_{23}\Pt_{12},\quad\qquad
D\Pe_{12}\Pt_{23}\Pt_{12}=\Pe_{23}\Pt_{12}, \label{useful1}  \\[8pt] 
D\: \Pt_{12}\Pt_{23}=\Pe_{23}\Pe_{12}\Pt_{23} ,\quad\qquad
D\: \Pe_{23}\Pt_{12}\Pt_{23}=\Pe_{12}\Pt_{23,} \label{useful1'}  \\[8pt] 
D\:\Pt_{23} \Pt_{12}=\Pt_{23}\Pe_{12}\Pe_{23} ,\quad\qquad
D\: \Pt_{23}\Pt_{12}\Pe_{23}=\Pt_{23}\Pe_{12}; \label{useful1''}  
\eea
Eq. (\ref{braid1}-\ref{useful1''}) hold also for   $l>3$, e.g. for all $2\le h\le l-1$
\be 
A_{(h-1)h}\,\Pe_{h(h+1)}\,\Pe_{(h-1)h}  
=\Pe_{h(h+1)}\,\Pe_{(h-1)h}\,A_{h(h+1)}. \label{braid2} 
\ee 
The {\it completely symmetric  trace-free} projectors
$\Ps^{l}$  generalize $\Ps^2\equiv\Ps$ to all $l> 2$.
$\Ps^{l}$  projects $\E^{\otimes^l}$ to the carrier space 
of the $l$-fold completely symmetric irrep  of $Uso(D)$, isomorphic to $\check V^{l}_D,V^{l}_D$,
therein contained. It is uniquely characterized by the following properties: \ for $n=1,...,l\!-\!1$,
\bea 
&&\Ps^{l}\Ps^-_{n(n\!+\!1)}=0,\qquad \Ps^-_{n(n\!+\!1)}\Ps^{l}=0,  \label{Plproj0} \\[6pt]  
&& \Ps^{l}\Pt_{n(n\!+\!1)}=0,\qquad \Pt_{n(n\!+\!1)}\Ps^{l}=0, \label{Plproj1}  \\[6pt]
&&\left(\Ps^{l}\right)^2=\Ps^{l},   \label{Plproj2}       
\eea
\begin{comment}
%Consequently, it is also \ $\mbox{tr}_{1\ldots l}\!\left({\cal P}^{l}\right) =\mbox{dim}(V^{l}_D)$, which guarantees that $\Ps^{l}$ acts as the identity (and not as a proper projector) on $V^{l}_D$. \ 
which imply also the ones
\be
 \Ps^{l}\Pe_{n(n\!+\!1)}=\Ps^{l}, \quad n=1,...,l\!-\!1;\qquad
\Ps^{l}\Ps^{h}_{(i+1)...(i+h)}=\Ps^{l},\quad  h<l, \:\: 0\le i\le l\!-\!h.
\label{furtherprop}
\ee 
\end{comment}
Eq.s (\ref{Plproj1}) amount to \ $\Ps^{l}{}^{i_1...i_l}_{j_1...j_l}\delta^{j_nj_{n+1}}=0$, \ $\delta_{i_ni_{n+1}}\Ps^{l}{}^{i_1...i_l}_{j_1...j_l}=0$. \
Proposition 3.2 of \cite{FioJPA23} yields a recursive construction of the projectors $\Ps^l$  (mimicking that of the quantum group $U_qso(D)$ covariant symmetric projectors of Proposition 1  of \cite{Fio04JPA}):  $\Ps^{l\!+\!1}$ can be expressed
as a polynomial in the permutators $\Pe_{12},...,\Pe_{(l\!-\!1)l}$  and trace projectors
$\Pt_{12},...,\Pt_{(l\!-\!1)l}$ through either recursive relation
\bea
\Ps^{l\!+\!1}&=&
\Ps^{l}_{12...l}M_{l(l\!+\!1)}\Ps^{l}_{12...l},
\label{ansatz1} \\[8pt]
&=& \Ps^{l}_{2...(l\!+\!1)}
M_{12}\Ps^{l}_{2...(l\!+\!1)}, \label{ansatz2} 
\eea
$M\equiv M(l\!+\!1) =\frac 1{l\!+\!1} \!\left[\1_{D^2}\!+\! l \Pe \!-\!
\frac{2Dl}{D\!+\!2l\!-\!2} \Pt\right]$.
%$$ \mbox{where } \quad M\equiv M(l\!+\!1) =\frac 1{l\!+\!1} \!\left[\1_{D^2}+ l\,\Pe -
%\frac{2Dl}{D\!+\!2l\!-\!2} \Pt\right];               \label{Ml}
%$$
All $\Ps^{l}$ are symmetric, $(\Ps^{l})^T=\Ps^{l}$. Let  
\be
X_l^{i_1...i_l}\equiv \Ps^l{}^{i_1...i_l}_{j_1...j_l} x^{j_1} ... x^{j_l},
%\in \check V^l_D,
\qquad  T_l^{i_1i_2...i_l}\equiv r^{-l}\, X_l^{i_1i_2...i_l}=\Ps^l{}^{i_1...i_l}_{j_1...j_l} t^{j_1} ... t^{j_l}. 
\label{defXTD}
\ee
Using  (\ref{Plproj1}) one easily shows that  $\Delta X_l^{i_1...i_l}=0$: the harmonic homogeneous $x^i$-polynomials $X_l^{i_1...i_l}$ make up a complete set of $\check V^l_D$ (not a basis, because they are invariant under 
permutations of $(i_1...i_l)$ and fulfill %the linear dependence relations
$\delta_{i_ni_{n+1}}X_l^{i_1...i_l}=0$, $n=1,...,l\!-\!1$). Similarly, the $t^i$-polynomials $T_l^{i_1...i_l}$
make up a complete set $\T_l$ (but not a basis) of $V_D^l$ that is easier to work with than the basis of spherical harmonics. Moreover, $\bL^2$,
$iL_{hk}$ and the multiplication operators $t^{h}\cdot$ act on the $T_l^{i_1...i_l}$ as follows:
\bea
&& \bL^2\,T_l^{i_1...i_l}  = E_l\,T_l^{i_1...i_l},
\label{LeigenvectorsT} \\[8pt]
&&\ba{lll}
iL_{hk}T_l^{i_1...i_l} &= &\displaystyle (l\!+\!1)\,
\frac {D\!+\!2l\!- \!2}{D\!+\!2l}
\left(\Ps^{l+1}{}^{hi_1...i_l}_{kj_1...j_l}
-\Ps^{l+1}{}^{ki_1...i_l}_{hj_1...j_l}\right)T_l^{j_1...j_l}, \\[8pt]
&= & \displaystyle l\,\Ps^l{}^{i_1...i_l}_{j_1...j_l}
\left(\delta^{kj_1}T_{l}^{hj_2...j_l}-\delta^{hj_1}T_{l}^{kj_2...j_l} \right),\ea    \label{LonT} 
\\[10pt]
&& t^{h}\,T_{l}^{i_1...i_l} = T_{l+1}^{hi_1...i_l}+\frac {l}{D\!+\!2l\!- \!2}\,
\Ps^{l}{}^{i_1i_2...i_l}_{hj_2...j_l} T_{l-1}^{j_2...j_l}\: \in\:   V_D^{l+1}\oplus   V_D^{l-1},   \label{tTdeco}
\\[6pt]
&& t^iT_l^{ii_2...i_l}=  \frac {1}{D\!+\!2l\!-\!2}
\left[D\!+\! l \!-\!1\!-\!
\frac{2l\!-\!2}{D\!+\!2l\!-\!4}\right]
T_{l-1}^{i_2...i_l}        \: \in\:      V_D^{l-1}.     \label{tTcontraction}
\eea
These formulae immediately follow from analogous ones for the  $X_l^{i_1...i_l}$.
More generally, the product $T_l^{i_1...i_l}T_m^{j_1...j_m}$ decomposes as follows
into  $V_D^n$ components:
\bea
T_l^{i_1...i_l}T_m^{j_1...j_m}= \sum\limits_{n\in \I^{lm}}\, 
S^{i_1...i_l,j_1...j_m}_{k_1...k_n}T_n^{k_1...k_n}, 
\label{TTdeco}
\eea
where  $\I^{lm}\equiv \{|l\!-\!m|,|l\!-\!m|\!+\!2,...,l\!+\!m\}$ and,  defining \ 
$\displaystyle s\!\equiv \!\frac{l\!+\!m\!-\!n}2\in\{0,1,...,m\}$, 
\bea
\ba{l}
\displaystyle
S^{i_1...i_l,j_1...j_m}_{k_1...k_n}= N^{lm}_n \,V^{i_1...i_l,j_1...j_m}_{k_1...k_n}, \qquad
N^{lm}_n = \frac{(D\!+\!2n\!-\!2)!!\,l!\,  m!}{(D\!+\!2n\!+\!2s\!-\!2)!!\, (l\!-\!s)!\,(m\!-\!s)!}\\[10pt]
V^{i_1...i_l,j_1...j_m}_{k_1...k_n} =  
 \Ps^l{}_{a_1...a_sc_{1}...c_{l-s}}^{i_1...i_l}\Ps^m{}_{a_1...a_s c_{l-s+1}...c_n}^{j_1...j_sj_{s+1}...j_m} \Ps^n{}_{c_1...c_n} ^{k_1...k_n}. 
\ea\label{Scondi''}
\eea
Thus the $S^{i_1...i_l,j_1...j_m}_{k_1...k_n}$ play the role of Clebsch-Gordon coefficients in the decomposition of a product of spherical harmonics. 
Finally,  \ $\big\la T_l^{i_1...i_l},T_n^{j_1...j_n}\big\ra \propto 
\delta_{ln} \, \Ps^l{}_{i_1...i_l}^{j_1...j_l}$ \ w.r.t. the scalar product of ${\cal L}^2(S^d)$. 

\subsection{Embedding in $\RR^{D+1}$,  isomorphism
$\mbox{End}\!\left(Pol_D^\Lambda\right)\simeq {\bf \pi}_{D+1}^{\Lambda}\big[Uso(D\!+\!1)\big]$}
\label{Embeddings}

Henceforth we abbreviate $\bD\equiv D+1$. We naturally embed $\CC[\RR^D]\hookrightarrow\CC[\RR^\bD]$; we use real Cartesian coordinates $(x^i)$ for $\RR^D$ and $(x^I)$ for $\RR^\bD$; $h,i,j,k\in\{1,...,D\}$, $H,I,J,K\in\{1,...,\bD\}$. We naturally embed $O(D)\hookrightarrow SO(\bD)$ identifying
$O(D)$ as the subgroup  of $SO(\bD)$ that is the little group of the  $\bD$-th axis; its Lie algebra, isomorphic to $so(D)$,  is 
generated by the $L_{hk}$. 
We shall add $\bD$ as a subscript to distinguish objects in dimension $\bD$ from their counterparts in dimension $D$, e.g. the distance $r_\bD$   from the origin in $\RR^\bD$, 
from its counterpart $r\equiv r_D$ in $\RR^D$, $\Ps^l_\bD$ from  $\Ps^l\equiv \Ps^l_D$, 
and so on. Setting  $t^I\equiv r_\bD^{-1}x^I$, for  $\Lambda\in\NN_0$ 
$\check V_\bD^\Lambda$, $V_\bD^\Lambda= r_\bD^{-\Lambda}\check V_\bD^\Lambda$
are respectively spanned by the 
\be
X_{\bD,\Lambda}^{I_1...I_\Lambda}=\Ps_\bD^\Lambda{}^{I_1...I_\Lambda}_{J_1...J_\Lambda} x^{J_1} ... x^{J_\Lambda},
%\in \check V^\Lambda_\bD,
\qquad T_{\bD,\Lambda}^{I_1...I_\Lambda}= r_\bD^{-\Lambda}X_{\bD,\Lambda}^{I_1...I_\Lambda} =\Ps_\bD^\Lambda{}^{I_1...I_\Lambda}_{J_1...J_\Lambda} t^{J_1} ... t^{J_\Lambda}.
%\in  V^l_\bD,
\label{defXTbD}
\ee 
The following combinations of the latter factorize into $X_{l}^{i_1...i_l}$ (resp. $T_{l}^{i_1...i_l}$) times a $O(D)$-scalar:
\bea
\check F^{i_1...i_l}_{\bD,\Lambda} \equiv   \Ps^l{}^{i_1...i_l}_{j_1...j_l}X_{\bD,\Lambda}^{j_1...j_l\bD...\bD} =
 \check p_{\Lambda,l}\,X_{l}^{i_1...i_l}, \qquad
F^{i_1...i_l}_{\bD,\Lambda}\equiv  r_\bD^{-\Lambda}\check F^{i_1...i_l}_{\bD,\Lambda}=
p_{\Lambda,l}T_{l}^{i_1...i_l}
\label{cFFDeco}
\eea
where $\check p_{\Lambda,l}$ is the homogeneous polynomial of degree \ $\Lambda-l$\  in  $x^\bD,r_\bD$ 
\bea
\check p_{\Lambda,l}=\left(x^\bD\right)^{\Lambda-l}+\left(x^\bD\right)^{\Lambda-l-2}r_\bD^2\, b_{\Lambda,l+2}
+\left(x^\bD\right)^{\Lambda-l-4} r_\bD^4\, b_{\Lambda,l+4}+...\:\: ,  \label{cp_h}\\[8pt]
b_{\Lambda,l+2k}=(-)^k\frac{(\Lambda\!-\!l)!\,(2\Lambda\!-\!4\!-\!2k\!+\! \bD)!!}{(\Lambda\!-\!l\!-\!2k)!\,(2k)!!\,(2\Lambda\!-\!4\!+\! \bD)!!}, \qquad k=1,2,....\left[\frac{\Lambda\!-\!l}{2}\right] , \label{bcoeff}
\eea
and  $p_{\Lambda,l}\equiv \check p_{\Lambda,l}(x^\bD,r_\bD)\, r_\bD^{l-\Lambda}$ is a polynomial of degree $h=\Lambda-l$ in $t^\bD$ only.
Hence the $F^{i_1...i_l}_{\bD,\Lambda}$ are eigenvectors of $\bL^2$ with eigenvalue $E_l$,  
 transform under $L_{hk}$ as the $T^{i_1...i_l}_l$ and under $L_{h\bD}$ as follows:
\bea
 iL_{h\bD} F^{i_1...i_l}_{\bD,\Lambda}=
 ( \Lambda\!-\!l)\, F_{\bD,\Lambda}^{hi_1...i_l}-\frac {l(\Lambda\!+\!l\!+\!D\!-\!2)}{D\!+\!2l\!-\!2}\: \Ps^l{}^{i_1i_2...i_l}_{hj_2...j_l}\, F_{\bD,\Lambda}^{j_2...j_l}\, .
\label{L-su-F}
\eea
These relations follow from exactly the same relations for the $\check F^{i_1...i_l}_{\bD,\Lambda}$.
As a consequence, 
$\check V_\bD^\Lambda$, $V_\bD^\Lambda$ decompose into irreducible components of $Uso(D)$ as follows:
\be
\check V_\bD^\Lambda=\bigoplus_{l=0}^\Lambda \check V_{D,\Lambda}^l,\qquad
V_\bD^\Lambda=\bigoplus_{l=0}^\Lambda V_{D,\Lambda}^l,
\ee
where \ $ \check V_{D,\Lambda}^l \!\simeq\! V_D^l$,  $V_{D,\Lambda}^l \!\simeq\! V_D^l$ \ are resp.  spanned by  the $ \check F^{i_1...i_l}_{\bD,\Lambda}$,
 $ F^{i_1...i_l}_{\bD,\Lambda}$. For $\Lambda=0,1,2$ we have: \ 
$\check V_\bD^0\simeq V_\bD^0\simeq\CC\simeq V_D^0$. \ $\check V_{D,1}^0,V_{D,1}^0$  are isomorphic to $V_D^0$ and resp. spanned by $x^\bD,t^\bD$; \  $\check V_{D,1}^1,V_{D,1}^1$  are isomorphic to $V_D^1$ and resp. spanned by the $x^i,t^i$. \
$\check V_{D,2}^0, V_{D,2}^0$  are isomorphic to $V_D^0$ and resp. spanned by
$X_{\bD,2}^{\bD\bD}=x^\bD x^\bD-r_\bD^2/\bD$, $F_{\bD,2}=T_{\bD,2}^{\bD\bD}=t^\bD t^\bD-1/\bD=D/\bD-\sum\nolimits_{h=0}^D t^ht^h$; \
$\check V_{D,2}^1,V_{D,2}^1$   are isomorphic to $V_D^0$ and resp. spanned by the
$\check F_{\bD,2}^i=X_{\bD,2}^{i\bD}=x^i x^\bD$, $F_{\bD,2}^i=T_{\bD,2}^{i\bD}=t^i t^\bD$; \ $\check V_{D,2}^2,V_{D,2}^2$   are isomorphic to $V_D^2$ and resp. spanned by the
$\check F_{D,2}^{ij}=X_{\bD,2}^{ij}+X_{\bD,2}^{\bD\bD}\delta^{ij}/D=X_{2}^{ij}$, 
$F_{D,2}^{ij} %\equiv \Ps^{ij}_{hk}T_{\bD,2}^{hk}
=T_{\bD,2}^{ij}+\frac{\delta^{ij}}{D}T_{\bD,2}^{\bD\bD}
%=T_{\bD,2}^{ij}-\frac{\delta^{ij}}{D}\sum\nolimits_{h=0}^D T_{\bD,2}^{hh}
%=t^it^j- \frac{\delta^{ij}}{D}\sum\nolimits_{h=0}^D t^ht^h
=T_{2}^{ij}$; 
the last equalities follow from $X_2^{ij}=x^ix^j\!-\!r^2 \frac{\delta^{ij}}D$,
$X_{\bD,2}^{ij}=x^ix^j\!-\!r_\bD^2\frac{\delta^{ij}}\bD$, $T_2^{ij}=t^it^j\!-\!\frac{\delta^{ij}}D$,  $T_{\bD,2}^{ij}=t^it^j\!-\!\frac{\delta^{ij}}\bD$.
%This shows
%the decomposition \ $V_\bD^2= V_{D,2}^2\oplus V_{D,2}^1\oplus V_{D,2}^0\simeq Pol_D^2$ \  explicitly. 

\section{Relations among the $\overline x^i,\overline L_{hk}$, isomorphisms  of $\Hi_{\Lambda},\A_{\Lambda}$,  $*$-automorphisms of $\A_{\Lambda}$}
\label{xLRel}

The  functions  $\bpsi_l^{i_1i_2...i_l}\!\equiv\! T_{l}^{i_1i_2...i_l}\! f_{l}$  with fixed $l$
 make up a complete set $\Ss^l_{D,\Lambda}$ in the eigenspace 
$\Hi_\Lambda^l$ of $H$, $\bL^2$  with eigenvalues $E_{0,l}$, $E_l$. 
$\Ss_{D,\Lambda}\equiv \cup_{l=0}^\Lambda\Ss^l_{D,\Lambda}$ is complete in $\Hi_\Lambda$. 
The $\overline L_{hk}, \overline{x}^i$ act as
\bea
i\overline L_{hk}\, \bpsi_l^{i_1i_2...i_l}
%&= &\displaystyle l\,\frac{\zeta_{l+2}}{\zeta_{l+1}}\left(\Ps^{l+1}{}^{hi_1...i_l}_{kj_1...j_l} -\Ps^{l+1}{}^{ki_1...i_l}_{hj_1...j_l}\right)\bpsi_l^{j_1...j_l}, \\[8pt]
\: = \: \displaystyle l\,\Ps^l{}^{i_1...i_l}_{j_1...j_l}
\left(\delta^{kj_1}\bpsi_{l}^{hj_2...j_l}-\delta^{hj_1}\bpsi_{l}^{kj_2...j_l} \right), 
  \label{Lonpsi}  \\[8pt]
\overline{x}^i\, \bpsi_l^{i_1i_2...i_l} \: = \: c_{l+1}\,\bpsi_{l+1}^{ii_1...i_l}+
\frac{c_{l}\, l}{D\!+\!2l\!-\!2}\,\Ps^{l}{}^{i_1i_2...i_l}_{ij_2...j_l}\, \bpsi_{l-1}^{j_2...j_l},\label{xonpsi} \\[4pt]
\mbox{where }\quad c_{l} \equiv   \left\{\!\!\ba{ll}
\sqrt{1+\frac{(2D\!-\!5)(D\!-1\!)}{2k}+\frac{(l\!-\!1)(l\!+\!D\!-\!2)}{k}} \:&\mbox{if }
1 \le l\le\Lambda, \\[6pt]
0 \:&%\mbox{if }l=0,\Lambda,
\mbox{otherwise.}\ea\right. \nonumber %\quad \label{defc}
\eea
Eq. (\ref{Lonpsi}) follows from  (\ref{LonT}), while (\ref{xonpsi}) holds up to  $O\left(k^{-3/2}\right)$ corrections that depend 
on the terms proportional to $(r\!-\!1)^k$, $k>2$, in the Taylor expansion of $V$ and 
could be made vanish by suitably choosing $V$. Henceforth we  adopt (\ref{Lonpsi}-\ref{xonpsi}) as exact {\it definitions} of  $\overline{L}_{hk},\overline{x}^i$. 
By Proposition 4.1 in \cite{FioJPA23}, 
the $\overline{L}_{hk},\overline{x}^i$ defined by (\ref{Lonpsi}-\ref{xonpsi}) are self-adjoint operators generating the 
$N^2$-dimensional $*$-algebra 
\ $\A_{\Lambda}\equiv End(\Hi_{\Lambda})\simeq M_{N}(\CC)$ \ of observables  on  $\Hi_{\Lambda}$; \ here $N\equiv \frac{(D\!+\!\Lambda\!-\!2)...(\Lambda\!+\!1)}{(D\!-\!1)!}(D\!+\!2\Lambda\!-\!1)$. Abbreviating \ $\overline{\bx}^2\equiv \overline{x}^i\,\overline{x}^i$, $\overline{\bL}^2\equiv \overline{L}_{ij}\,\overline{L}_{ij}/2$, $B\equiv (2D\!-\!5)(D\!-1\!)/2$, \ they  fulfill the relations
\bea
&& \big[i\overline{L}_{ij},\overline{x}^h\big]= \overline{x}^i\delta^h_j\!-\!\overline{x}^j\delta^h_i ,\label{linea1}\\[8pt]
&& \big[i\overline{L}_{ij},i\overline{L}_{hk}\big]=i\left(\overline{L}_{ik}\delta^j_h\!-\!\overline{L}_{jk}\delta^i_h\!-\!\overline{L}_{ih}\delta^j_k\!+\!\overline{L}_{jh}\delta^i_k\right), 
\label{linea4}\\[8pt]
&& \varepsilon^{i_1i_2i_3....i_D}\overline{x}^{i_1}\overline{L}_{i_2i_3} =0,  \qquad D\ge 3,  \label{linea3}\\[8pt]
&& (\overline{x}^h\!\pm\!i\overline{x}^k)^{2\Lambda+1}=0, \quad
(\bar L^{hj}\!+\!i\bar L^{kj})^{2\Lambda+1}=0,   \qquad
\mbox{if }h\neq j\neq k\neq h, \label{linea5}\\[8pt]
&&  \left[\overline{x}^i,\overline{x}^j\right] = i\overline{L}_{ij} \left(-\frac{I}{k}\!+\!K\,P_{\Lambda}^{\Lambda}\right), \quad 
\ba{l} K\equiv 
%\frac{1}{k}+\frac{(\rho_{\Lambda-1,\Lambda})^2}{D\!+\!2\Lambda\!- \!2}=
\frac{1}{k}\!+\!\frac{1}{D\!+\!2\Lambda\!- \!2}\left[1\!+\!\frac{B}{k}\!+\!
\frac{(\Lambda\!-\!1)(\Lambda\!+\!D\!-\!2)}{k}\right],\ea
\label{linea2}\\[8pt]
&& \overline{\bx}^2=1\!+\!\frac {\overline{\bL}^2}k \!+\!\frac{B}{k}
-\frac{\Lambda\!+\!D\!-\!2}{2\Lambda\!+\!D\!-\!2}
\left[1\!+\!\frac{B}{k}\!+\!\frac{\Lambda (\Lambda\!+\!D\!-\!1)}{k}\right]  P_{\Lambda}^{\Lambda}  
=:\chi(\bL^2). \label{bx^2}
\eea
A fuzzy sphere is obtained choosing $k$ as a function $k(\Lambda)$ fulfilling (\ref{consistencyD}), e.g. \ $k=\Lambda^2(\Lambda\!+\!D\!-\!2)^2/4$; 	\ the commutative limit is \
$\Lambda\rightarrow\infty$. We remark that:

\begin{enumerate}[label=\thesection.\alph*]

\item Eq. (\ref{linea3}) is the analog of (\ref{Lijrel}b).
By (\ref{linea2}), it  can be reformulated also as $\varepsilon^{i_1i_2i_3....i_D}\overline{x}^{i_1}\overline{x}^{i_2}\overline{x}^{i_3} =0$.

\item  By   (\ref{bx^2}), (\ref{Proj-l})$_{l=\Lambda}$ \ $\overline{\bx}^2$
is not a constant, but  can be expressed as a polynomial $\chi$ in $\overline{\bL}^2$ only, with the same eigenspaces $\Hi_{\Lambda}^l$. All its eigenvalues $r^2_l$, except $r^2_\Lambda$,
are close to 1, slightly (but strictly) grow with $l$ and collapse to 1 as $\Lambda\to \infty$. Conversely,
$\overline{\bL}^2$ can be expressed as a polynomial $\upsilon$ in $\overline{\bx}^2$, via \
$\overline{\bL}^2=\sum_{l=0}^\Lambda E_l  P^l_{\Lambda}$ and 
$ P^l_{\Lambda}=\prod_{n=0, n\neq l}^{\Lambda} \frac{\overline{\bx}^2-r^2_n}
{r^2_l-r^2_n}$.

\item By (\ref{linea2}),  (\ref{Proj-l})$_{l=\Lambda}$ the commutators $[\overline{x}^i,\overline{x}^j]$  are Snyder-like, i.e. of the form $\alpha \overline{L}_{ij}$; also $\alpha$ depends only on the $\overline{L}_{hk}$,
more precisely can be expressed as a polynomial in $\overline{\bL}^2$.
%, and vanish  as \ $\Lambda\!\to\! \infty$.    

\item  Using (\ref{linea1}), (\ref{linea4}), (\ref{linea2}), all polynomials in $\overline{x}^i,\overline{L}_{hk}$  can be expressed as combinations of  monomials in $\overline{x}^i,\overline{L}_{hk}$ in any prescribed order, e.g. in the natural one
\be
\big(\overline{x}^1\big)^{n_1}...\big(\overline{x}^D\big)^{n_D}
\big(\overline{L}_{12}\big)^{n_{12}}\big(\overline{L}_{13}\big)^{n_{13}}...\big(\overline{L}_{dD}\big)^{n_{dD}}, \qquad n_i,n_{ij}\in\NN_0; \label{monomials}
\ee
 the coefficients, which can be put at the right of these monomials, are complex combinations of 1 and $P^{\Lambda}_{\Lambda}$.   Also $P^{\Lambda}_{\Lambda}$ can be expressed as a polynomial in $\overline{\bL}^2$ via (\ref{Proj-l})$_{l=\Lambda}$.
%[their degrees are bounded by (\ref{rf3D3}-\ref{rf3D4})] 
Hence a suitable subset of such ordered monomials makes up a   basis of the $N^2$-dim  vector space \ $\A_\Lambda$.

\item Actually,  $\overline{x}^i$ {\it generate} the $*$-algebra $\A_\Lambda$, because 
also the $\overline{L}_{ij}$ can be expressed as {\it non-ordered} polynomials in the $\overline{x}^i$: 
by  (\ref{linea2}) $\overline{L}_{ij}=[\overline{x}^j,\overline{x}^i]/\alpha$, and also $1/\alpha$, which depends only on $  P^{\Lambda}_{\Lambda}$, can be expressed itself as a polynomial in $\overline{\bx}^2$, as shown above.

\item  Eq. (\ref{linea1}-\ref{bx^2})  are equivariant under the whole group $O(D)$, including
the inversion  $\overline{x}^i\!\mapsto\!-\overline{x}^i$
%$\overline{L}_i\!\mapsto\!\overline{L}_i$
of one axis, or more  (e.g. parity), contrary to Madore's and Hoppe's FS.

\end{enumerate}
We slightly enlarge $Uso(D)$ by introducing the new generator 
 $\lambda= \left[\sqrt{(D\!-\!2)^2+4\bL^2}-D+2\right]/2$,
which fulfills \ $\lambda(\lambda\!+\!D\!-\!2)=\bL^2$, \ so that $V_D^l$ is a $\lambda=l$ eigenspace, and \ $\lambda \, F^{i_1...i_l}_{\bD,\Lambda}
=l \,F^{i_1...i_l}_{\bD,\Lambda}$. Theorem 5.1 in \cite{FioJPA23} states that there exist a $O(D)$-module isomorphism 
$\varkappa_\Lambda:\Hi_\Lambda\rightarrow V_\bD^\Lambda$ and a $O(D)$-equivariant
algebra  map  $\kappa_\Lambda:\A_\Lambda\equiv\mbox{End}\!\left(\Hi_\Lambda\right)\to \bpi_\bD^\Lambda\big[Uso(\bD)\big]$, $\bD\equiv D\!+\!1$, such that
\be
\varkappa_\Lambda(a\bpsi)=\kappa_\Lambda(a)\varkappa_\Lambda(\bpsi), \qquad\forall\:\bpsi
\in \Hi_\Lambda,
\quad  a\in\A_\Lambda\:.
\label{compatibilityCond'}
\ee
On the  $\bpsi_{l}^{i_1...i_l}$ (spanning $\Hi_\Lambda$) and on generators \ $L_{hi},\, \overline{x}^i$ \ of $\A_\Lambda$ they respectively act as follows:
\bea
&& \varkappa_\Lambda\big(\bpsi_{l}^{i_1...i_l}\big)\equiv  a_{\Lambda,l}F^{i_1...i_l}_{\bD,\Lambda}=a_{\Lambda,l} \, p_{\Lambda,l}\, T_{l}^{i_1...i_l},\qquad l=0,1,...,\Lambda ,       \label{Tcorr'}\\[8pt]
&&  \kappa_\Lambda\big(\overline{L}_{hi}\big)\equiv  \bpi_\bD^\Lambda (L_{h i})\,,\qquad \kappa_\Lambda\big(\overline{x}^i\big)\equiv  \bpi_\bD^\Lambda\big[m_\Lambda^*(\lambda) \, X^i\, m_\Lambda(\lambda)\big]\,,                                  \label{Opcorr'}
\eea
where \ $X^i\equiv L_{\bD i}$, \ $A\equiv \sqrt{k+(D\!-\!1)(D\!-\!3)3/4}$, \   $\Gamma$ is Euler gamma function, and
\bea
&& a_{\Lambda,l}=a_{\Lambda,0}\,i^l \sqrt{\frac{\Lambda(\Lambda\!-\!1)...(\Lambda\!-\!l\!+\!1)}{(\Lambda\!+\!D\!-\!1)(\Lambda\!+\!D)...(\Lambda\!+\!l\!+\!D\!-\!2)}}, \label{mO(D)''}\\[10pt]
&&  m_{\Lambda}(s)=
%\frac 12 \sqrt{\frac{\Gamma\!\left(\frac {\Lambda\!+\!s\!+\!D\!-\!1}2\right)\,\Gamma\!\left(\frac {\Lambda\!-\!s\!+\!1}2\right)}{\Gamma\!\left(\frac {\Lambda\!+\!s\!+\!D}2\right)\, \Gamma\!\left(\frac {\Lambda\!-\!s}2\!+\!1\right)}}
\sqrt{\frac{\Gamma\!\left(\frac {\Lambda+s+d}2\right)\,\Gamma\!\left(\frac {\Lambda-s+1}2\right)\,\Gamma\!\left(\frac {s+1+d/2+iA}2\right)\,\Gamma\!\left(\frac {s+1+d/2-iA}2\right)}{\Gamma\!\left(\frac {\Lambda+s+D}2\right)\, \Gamma\!\left(\frac {\Lambda-s}2+1\right)\,\Gamma\!\left(\frac {s+d/2+iA}2\right)\,\Gamma\!\left(\frac {s+d/2-iA}2\right)\,\sqrt{k}}} .\label{mO(D)'}
\eea

\noindent
Finally,   $*$-automorphisms $\omega$ of $\A_{\Lambda}\simeq M_N(\CC)$ are inner and make up
a group $G\simeq SU(N)$, i.e. 
\be
\omega\, : \, a\in  M_N(\CC)\mapsto g\, a \, g^{-1}\in  M_N(\CC)    \label{autom}
\ee
for some unitary $N\times N$ matrix $g$  with $\det g=1$.  Consider the $G$-subgroup \ $G'\equiv\{g=\bpi_\bD^\Lambda\left[e^{i\alpha}\right]\, |\,\alpha\in so(\bD)\}\simeq SO(\bD)$. \
Choosing %$\alpha=\alpha^{ij}L_{ij}$ ($\alpha^{ij}\in\RR$)  
$\alpha\in so(D)\subset so(\bD)$ the automorphism amounts to 
  a $SO(D)\subset SO(\bD)$ transformation, i.e. a rotation in the $x\equiv(x^1,...,x^D)\in\RR^D$ space. The $O(D)\subset SO(\bD)$ transformations with determinant  $-1$ 
keep the same form also in the $\overline  X\equiv (X^1,...,X^D)$ 
and [by (\ref{Opcorr'})] in the $\overline  x\equiv (\overline x^1,...,\overline x^D)$ spaces.
In particular, those inverting one or more axes of $\RR^D$ (i.e. changing the sign
of one or more $x^i$, and thus also of $X^i,\overline  x^i$), e.g. parity,
can be also realized as  $SO(\bD)$ transformations, i.e. rotations in $\RR^\bD$.
This shows that (\ref{Opcorr'}) is equivariant under the whole
$O(D)$, which plays the role of isometry group of this fuzzy sphere.

\section{Fuzzy spherical harmonics, and limit $\Lambda\to\infty$}
\label{Dconverge}

It's simpler to work with the $T_l^{i_1...i_l}$ than spherical harmonics, their combinations. In $\Hi_s={\cal L}^2(S^d)$ we have $\bpsi_l^{i_1...i_l}\propto T_l^{i_1...i_l}$, $\bpsi_0\propto 1 $. The $T_l^{i_1...i_l}\in C(S^d)$ act on $\Hi_s$ as multiplication operators  %trivially 
fulfilling $T_l^{i_1...i_l}\cdot \bpsi_0\propto \bpsi_l^{i_1...i_l}$.  We define their  $\Lambda$-th fuzzy analogs %$\widehat{T}_l^{i_1...i_l}$ 
  replacing \ $t^i\cdot\mapsto  \overline{x}^i$ \ in (\ref{defXTD}b), i.e. %we define 
\be
 \widehat{T}_l^{i_1...i_l} \equiv %\sum_{j_1,....,j_l}
\Ps^l{}^{i_1...i_l}_{j_1...j_l} \overline{x}^{j_1} ... \overline{x}^{j_l}, \qquad
\Rightarrow   \qquad    \widehat{T}_l^{i_1...i_l} \bpsi_0\propto \bpsi_l^{i_1...i_l}         \label{defhatT}
\ee
for $l\le\Lambda$. 
 Since $\bpsi_0$ is a scalar,  $\bpsi_l^{i_1...i_l},\widehat{T}_l^{i_1...i_l},T_l^{i_1...i_l}$ transform under $O(D)$ exactly in the same way, consistently with $\Hi_\Lambda\simeq Pol_D^{\Lambda}$.
As $\Lambda\to\infty$ the decomposition of $\Hi_\Lambda\simeq Pol_D^{\Lambda}$
into irreducible components under $O(D)$ 
 becomes isomorphic to the decomposition of $ \Hi_s \simeq Pol_D$. 
We define the $O(D)$-equivariant embedding \   ${\cal I}:\Hi_\Lambda\hookrightarrow \Hi_s$ \   by  setting \  ${\cal I}\big(\bpsi_l^{i_1...i_l}\big)\equiv T_l^{i_1...i_l}$ \
and applying the linear extension. Below we drop $\I$  and identify $\bpsi_l^{i_1...i_l}=T_l^{i_1...i_l}$  as elements of the Hilbert space $\Hi_s$. 
For all $\bphi\equiv\sum_{l=0}^{\infty}\phi^l_{i_1...i_l}T_l^{i_1...i_l}\in {\cal L}^2(S^2)$ and $\Lambda\in\NN$ let 
$\bphi_ \Lambda\equiv P_\Lambda\bphi=\sum_{l=0}^{\Lambda}\phi^l_{i_1...i_l}T_l^{i_1...i_l}$ be its projection to $\Hi_\Lambda$ (or $\Lambda$-th truncation). Clearly $\bphi_ \Lambda\to\bphi$
in the $\Hi_s$-norm $\Vert\,\Vert$:
in this simplified notation, \ $\Hi_\Lambda$ `invades' $\Hi_s$ as $\Lambda\to\infty$. 
${\cal I}$ induces the $O(D)$-equivariant embedding of operator algebras \ 
${\cal J}\!:\!\A_\Lambda\!\hookrightarrow\! B\left(\Hi_s\right)$ \
by setting\ $\J(a)\,\I(\bpsi)\equiv \I(a\bpsi)$;  \ here
$B\left(\Hi_s\right)$ stands for the $*$-algebra of bounded operators on $\Hi_s$.
By construction, $\A_\Lambda$ annihilates $\Hi_\Lambda^\perp$. \ 
In particular,  $\J\!\left(\overline{L}_{hk}\right)=L_{hk}P^{\Lambda}$, and
 $\overline{L}_{hk}\bphi\stackrel{\Lambda\to\infty}{\longrightarrow} L_{hk}\bphi$ \  
 for all $\bphi\!\in\! D(L_{hk})\equiv$ the domain of $L_{hk}$.
More generally, $f(\overline{L}_{hk})\to f(L_{hk})$ strongly on $D[f(L_{hk})]\subset\Hi_s$, for all measurable functions $f(s)$.
Continuous functions $f$ on $S^d$, acting as multiplication operators 
$f\cdot:\bphi\in\Hi_s\mapsto f\bphi\in\Hi_s$, make up a subalgebra  $C(S^d)$
of $B\left(\Hi_s\right)$. \
Clearly, $f$ belongs also to  $\Hi_s$. Since $Pol_D$ is dense in both $\Hi_s$,   $C(S^d)$,  
$f_N$ converges to $f$  as $N\to \infty$  in both the $\Hi_s$  and  the $C(S^d)$  norm.
Identifying \ $\bpsi_{l}^{i_1...i_l} \equiv T_{l}^{i_1...i_l}$, 
eq. (\ref{tTdeco}), (\ref{xonpsi}) become
\bea
&& t^{h}\,T_{l}^{i_1...i_l} = T_{l+1}^{hi_1...i_l}+d_l\,
\Ps^{l}{}^{i_1i_2...i_l}_{hj_2...j_l} T_{l-1}^{j_2...j_l}, \qquad  
d_l\equiv \frac {l}{D\!+\!2l\!- \!2}\label{tTdeco'}\\[6pt]
&& \overline{x}^hT_l^{i_1i_2...i_l}=c_{l+1}\,T_{l+1}^{hi_1...i_l}+
c_{l}\: d_l\,\Ps^{l}{}^{i_1i_2...i_l}_{hj_2...j_l}\, T_{l-1}^{j_2...j_l}.
\label{xonT}
\eea
Theorem 6.1 in \cite{FioJPA23} states that the action of the $\widehat{T}_l^{i_1...i_l}$ on $\Hi_\Lambda$ is determined by 
\bea
\widehat{T}_l^{i_1...i_l}T_m^{j_1...j_m}=\sum\nolimits_{n\in L}\,\widehat{N}^{lm}_n\,\Ps^l{}_{a_1...a_rc_{1}...c_{l-r}}^{i_1...i_l}\Ps^m{}_{a_1...a_r c_{l-r+1}...c_n}^{j_1...j_rj_{r+1}...j_m} \Ps^n{}_{c_1...c_n} ^{k_1...k_n}\, T_n^{k_1...k_n},
\label{hTTdeco}
\eea
with suitable coefficients $\widehat{N}^{lm}_n$, cf.  (\ref{TTdeco}-\ref{Scondi''}). 
%related to their classical limits $ N^{lm}_n>0$ of formula (\ref{TTdeco}) by
%\be
%\widehat{N}^{lm}_n=0\quad\mbox{{\rm if} }\:\:l\!-\!m>\Lambda,\qquad\quad
%N^{lm}_n\le \widehat{N}^{lm}_n\le N^{lm}_n (c_\Lambda)^l
%\quad\mbox{{\rm otherwise}}.      \label{hN} \ee
As a fuzzy analog  of the vector space $C(S^d)$ we adopt
\be
{\cal C}_\Lambda\equiv \left\{\hat f_{2\Lambda}\equiv \sum\nolimits_{l=0}^{2\Lambda}
%\sum_{i_1,....,i_l}
f^l_{i_1...i_l}  \widehat{T}_l^{i_1...i_l}\:|\: f^l_{i_1...i_l}\in\CC\right\}
\subset\A_\Lambda\subset B\big(\Hi_s\big);
\label{def_CLambdaD}
\ee
here the highest $l$ is $2\Lambda$ because the $\widehat{T}_l^{i_1...i_l}$ annihilate $\Hi_\Lambda$ if $l>2\Lambda$.  By construction,
\be
{\cal C}_\Lambda=\bigoplus\nolimits_{l=0}^{2\Lambda}  \widehat{V}_D^l,\qquad\qquad  \widehat{V}_D^l\equiv \left\{%\sum_{i_1,....,i_l}
f^l_{i_1...i_l}  \widehat{T}_l^{i_1...i_l}\:,\: f^l_{i_1...i_l}\in\CC\right\} \label{deco2}
\ee
is the decomposition
of ${\cal C}_\Lambda$ into irreducible components under $O(D)$. $\widehat{V}_D^l$ is trace-free
for all $l>0$. %, i.e. its projection on the singlet component $\widehat{V}_D^0$ is zero.  
In the limit $\Lambda\to\infty$ (\ref{deco2}) becomes the decomposition of $C(S^d)$. 
As a fuzzy analog  of $f\in C(S^d)$
we adopt the sum \ $\hat f_{2\Lambda}$ \ appearing in (\ref{def_CLambdaD})
with the coefficients of the expansion $f=\sum_{l=0}^{\infty}\sum_{i_1,....,i_l}f^l_{i_1...i_l} T_l^{i_1...i_l} $ up to $l=2\Lambda$.
Theorem 6.2 in \cite{FioJPA23} states that 
for all $f,g\in C(S^d)$ the following strong $\Lambda\rightarrow \infty$ limits hold: $\hat{f}_{2\Lambda}\rightarrow f\cdot,\widehat{\left(fg\right)}_{2\Lambda}\rightarrow fg$ and $\hat{f}_{2\Lambda}\hat{g}_{2\Lambda}\rightarrow fg\cdot$. 
However $\hat f_{2\Lambda}$  {\it does not} converge  to $f$ {\it in operator norm},
because the operator $\hat f_{2\Lambda}$ (a polynomial in the $\overline{x}^i$)   annihilates $\Hi_{\Lambda}^\perp$ (the orthogonal complement of $\Hi_{\Lambda}$),  since so do
the  $\overline{x}^i=P^\Lambda x^i\cdot P^\Lambda$.

\section{Discussion and conclusions}
\label{discuss}

We have obtained a sequence $\left\{(\Hi_\Lambda,\A_\Lambda)\right\}_{\Lambda\in\NN}$ of $O(D)$-equivariant approximations of quantum mechanics of a particle on  $S^d$; $\Hi_\Lambda$ is the  Hilbert space of states, $\A_\Lambda\!\equiv$End$(\Hi_\Lambda)$ is the associated  $*$-algebra of observables, $H_\Lambda\in\A_\Lambda$ is the free Hamiltonian (this may be modifed by adding interaction terms $H_I\in\A_\Lambda$, so that the new Hamiltonian still maps $\Hi_\Lambda$  into iself).
%The commutators  of the Cartesian coordinates $\overline x^i\!\in\!\A_\Lambda$  are of Snyder type, i.e. proportional to the angular momentum components $\overline L_{ij}$.
$\A_\Lambda$ is spanned by ordered monomials (\ref{monomials}) in $\overline{x}^i,\overline{L}_{ij}$  (of appropriately bounded degrees), in the same way as 
 the algebra $\A_s$ of observables on $\Hi_s$  is spanned by ordered monomials in $t^i,L_{ij}$. 
However, while $\overline x^i$ generate the whole $\A_\Lambda$ because $[\overline x^i,\overline x^j]\propto\overline L_{ij}$ (as in Snyder spaces \cite{Snyder}), this has no analog in $\A_s$, because $[t^i,t^j]=0$.
The square distance $\overline{\bx}^2$ from the origin is not 1, but
a function of $\bL^2$ with a spectrum very close to 1, collapsing to 1 as $\Lambda\to\infty$. Each pair $(\Hi_\Lambda,\A_\Lambda)$ is isomorphic
 to $\left(V^\Lambda_{\bD},\bpi_\Lambda[Uso(\bD)]\right)$, $\bD\!\equiv\!D\!+\!1$, also as $O(D)$-modules; \ $\bpi_\Lambda$ is the irrep of $Uso(\bD)$ on the space 
$V^\Lambda_{\bD}$ of harmonic polynomials of degree $\Lambda$ on $\RR^{\bD}$,
restricted to $S^D$. 
We have also described (section \ref{Dconverge}) the subspace $\C_{\Lambda}\subset\A_{\Lambda}$ of completely symmetrized trace-free  polynomials in the $\overline x^i$; this is also spanned by the
fuzzy analogs of spherical harmonics.
$\Hi_\Lambda,\A_\Lambda,\C_{\Lambda}$ carry reducible representations of $O(D)$; 
as $\Lambda\to\infty$ their decompositions into irreps respectively go to the decompositions of 
$\Hi_s\equiv {\cal L}^2(S^d)$,  of $\A_s$ and of $C(S^d)\subset\A_s$ (the continuous functions on $S^d$ act  on $\Hi_s$ as multiplication operators).
There are natural embeddings \ $\Hi_\Lambda\hookrightarrow \Hi_s$, 
$\C_\Lambda\hookrightarrow C(S^d)$ and $\A_\Lambda\hookrightarrow \A_s$ such that
$\Hi_\Lambda\to \Hi_s$ in the norm of $\Hi_s$, while
$\C_\Lambda\to C(S^d)$,  $\A_\Lambda\to \A_s$ strongly  as $\Lambda\to\infty$.

Reintroducing the physical angular momentum components
$l_{ij}\equiv \hbar L_{ij}$,  then in the $\hbar\to 0$ limit $\A_s$ endowed with 
the usual quantum Poisson bracket \ $\{f,g\}=[f,g]/i\hbar$ \
goes to the (commutative) Poisson algebra $\F$ of (polynomial) functions on the classical phase space $T^*S^d$, generated by $t^i,l_{ij}$. We can directly obtain $\F$ from $\A_\Lambda$  adopting a suitable $\Lambda$-dependent $\hbar$ going to zero as 
$\Lambda\to\infty$\footnote{It suffices that $\hbar(\Lambda)k(\Lambda)$ diverges; if e.g. $k=\Lambda^2(\Lambda\!+\!D\!-\!2)^2/4$, then $\hbar(\Lambda)=O(\Lambda^{-\alpha})$ with $0<\alpha<4$ is enough.}.  
More formally, we can regard $\{\A_\Lambda\}_{\Lambda\in\NN}$ as a fuzzy
quantization of a coadjoint orbit of $O(\bD)$ that goes to the classical phase space $T^*S^d$.
We recall that coadjoint orbits $\mathcal {O}_{\blambda }=\mathrm {Ad} _{G}^{*}\blambda $ of a Lie group $G$  are orbits  of the coadjoint action $\mathrm {Ad} _{G}^{*} $ inside 
the dual space $\mathfrak {g}^{*}$ of 
the Lie algebra $\mathfrak {g}$ of $G$ passing through $\blambda\in\mathfrak {g}^{*}$,
or equivalently homogeneous spaces $G/G_{\blambda }$, where $G_{\blambda }$ is the stabilizer of $\blambda$ w.r.t.  $\mathrm {Ad} _{G}^{*} $. They have a natural symplectic structure. If $G$ is compact semisimple, identifying  $\mathfrak{g}^*\simeq\mathfrak{g}$ via the (nondegenerate) Killing form, we can resp. rewrite %more explicitly 
these definitions  in the form 
\bea
\mathcal {O}_{\blambda }\equiv \left\{g \blambda g^{-1}\:|\: g\in G\right\}\subset\mathfrak{g}^* ,\qquad\mathcal {O}_{\blambda }\equiv  G/G_{\blambda } \quad \mbox{where }\: G_{\blambda }\equiv \left\{ g\in G \:|\: g \blambda g^{-1}= \blambda\right\}.
\eea
Clearly,  $G_{\Lambda\blambda }=G_{\blambda }$ for all $\Lambda\in \CC\setminus \{0\}$. Denoting as $\Hi_{\blambda}$ the (necessarily finite-dimensional) 
carrier space of the irrep with  highest weight $\blambda$, one can regard 
(see e.g. \cite{Haw99}) the sequence of
$\{\A_\Lambda\}_{\Lambda\in\NN}$, with
$\A_\Lambda\equiv \mbox{End}\left(\Hi_{\Lambda\blambda}\right)$,  
as a fuzzy quantization of the symplectic space $\mathcal {O}_{\blambda }\simeq G/G_{\blambda }$.
The Killing form $B$ of $so(\bD)$ gives \ 
$B(L_{HI},L_{JK})=2(\bD\!-\!2)\left(\delta^H_J\delta^I_K-\delta^H_K\delta^I_J\right)$ \
for all $H,I,J,K\in\{1,2,...,\bD\}$.
%Let  $\sigma\equiv \left[\frac{\bD}2\right]=$ rank of $so(\bD)$. 
As a basis of the Cartan subalgebra $\mathfrak{h}$ of $so(\bD)$ we pick $\{H_a\}_{a=1}^\sigma$, where $\sigma\equiv \left[\frac{\bD}2\right]=$ rank of $so(\bD)$,
\bea
H_\sigma\equiv L_{D\bD},\quad H_{\sigma-1}\equiv L_{(d-1)d},\quad ...,\quad H_1=\left\{\!\!\ba{ll} 
L_{12}\:\: &\mbox{if }\bD=2\sigma, \\[2pt]
L_{23}\:\:  &\mbox{if }\bD=2\sigma\!+\!1.\ea\right.
\eea
We choose the irrep of $Uso(\bD)$ on $ V^\Lambda_\bD\simeq\Hi_\Lambda$ and $\Omega^\Lambda_\bD\equiv (t^D\!+\!it^\bD)^\Lambda\in V^\Lambda_\bD$ as the highest weight vector.
The joint spectrum $\bLambda=(0,...,0,\Lambda)$ of $H\equiv (H_1,...,H_\sigma)$ is 
the weight associated to the $\mathfrak{h}$-basis.
Identifying  $\blambda\in \mathfrak{h}^*$  
with   $H_\blambda\in \mathfrak{h}$ via the Killing form, we
find that $H_\bLambda \propto H_\sigma=L_{D\bD}$.
The stabilizer of $H_\bLambda$ in $SO(\bD)$ is
$SO(2)\!\times\!  SO(d)$, where  $so(2),so(d)$ are
resp. spanned by $H_\bLambda$, the $L_{ij}$ with $i,j<D$.
Thus the coadjoint orbit $\mathcal {O}_{\bLambda}=SO(\bD)/\big(SO(2)\!\times\! SO(d)\big)$
has the dimension of $T^*S^{d}$, 
%the cotangent space of the $d$-dimensional sphere $S^d$ (or phase space over $S^d$), 
$$
\ba{l}\frac{D(D+1)}2-1-\frac{(D-2)(D-1)}2=2(D-1)=2d, \ea
$$
consistently with the interpretation of 
$\A_\Lambda$ as the algebra of observables (quantized phase space)  on the fuzzy sphere.
It would have not been the case with some other irrep of
$Uso(\bD)$; $\mathcal {O}_{\blambda }$ would have been some other equivariant bundle over 
$S^d$ \cite{Haw99}.
For instance, the fuzzy spheres of dimension $d>2$ of \cite{GroKliPre96,Ramgoolam,DolOCon03,DolOConPre03} 
are based on $End(V^\Lambda)$, 
where the spaces $V^\Lambda$ carry {\it irreps} of both $Spin(D)$ and $Spin(\bD)$, hence of both $Uso(D)$ and  $Uso(\bD)$. Then: \ i) \ for some $\Lambda$ these may be only {\it projective} representations of
$O(D)$;  \ ii) \ in general (\ref{linea3}) will not be satisfied; \ iii) \
as $\Lambda\to \infty$ $V^\Lambda$ does not go to ${\cal L}^2(S^d)$ as a representation of $Uso(D)$, in contrast with our $\Hi_\Lambda\simeq V^\Lambda_{\bD}$; \ iv) \
the central $\bx^2\equiv X^iX^i$ can be normalized to $\bx^2=1$. Here $L_{i\bD}$ play the role of fuzzy coordinates $X^i$. In \cite{Ste16,Ste17}  $d=4$ and
$\mathcal {O}_{\blambda }=\CC P^3$, which has dimension 6 and can be seen as a $so(5)$-equivariant $S^2$ bundle over $S^4$.
Ref. \cite{Ste16,Ste17} constructs also a fuzzy 4-sphere $S^4_N$ 
based on based on a sequence of $End(V)$, 
where each $V$ carries an irrep $\pi$ of $Uso(6)$ which splits into the direct sum
of a small number $m>1$ of irreps of $Uso(5)$; the $O(5)$-scalar $\bx^2=X^iX^i$ is no longer central, but its spectrum is still very close to 1  provided.  The associated coadjoint orbit is 10-dimensional  and can be seen as a $so(5)$-equivariant $\CC P^2$ bundle over $\CC P^3$, or a $so(5)$-equivariant twisted bundle over either $S^4_N$ or $S^4_n$. 

\smallskip

$\A_s$ is generated by all the $t^h,L_{ij}$ with $h\!\le\! D$, $i\!<\! j\!\le\!  D$ (subject to the relations $t^it^h=t^ht^i$, $t^it^i=1$, $[iL_{ij},t^h]= t^i\delta^h_j-t^j\delta^h_i$, etc.), and $C(S^d)$ is generated  by the  $t^h$ alone. On the contrary, by Remark \ref{xLRel}.e
the  $\overline{x}^i$ {\it alone} generate 
the whole $\A_\Lambda\simeq\bpi_\bD^\Lambda\big[Uso(\bD)\big]$, which contains 
${\cal C}_\Lambda$ as a {\it proper} subspace, albeit not as a subalgebra; also the simpler generators $X^i=L_{\bD i}$ alone generate  $\A_\Lambda\simeq\bpi_\bD^\Lambda\big[Uso(\bD)\big]$, because of $L_{ij}=i[X^j,X^i]$ and (\ref{Opcorr'}).
Thus the Hilbert-Poincaré series of the algebra  generated by the 
$\overline{x}^i$ (or $X^i$), $\A_\Lambda$, is larger than that of   $Pol_D^\Lambda$
and  ${\cal C}_\Lambda$.
If by a ``quantized space"  we understand
a noncommutative deformation of the {\it algebra} of functions on that space
{\it preserving the Hilbert-Poincaré series}, then $\{\A_\Lambda\}_{\Lambda\in\NN}$  is a 
($O(D)$-equivariant, fuzzy) quantization of $T^*S^d$, the  phase space on $S^d$, while
$\{{\cal C}_\Lambda\}_{\Lambda\in\NN}$  is not a quantization of $S^d$, nor are the other fuzzy spheres,
except the Madore-Hoppe fuzzy 2-dimensional sphere: all the others, as ours, have the same Hilbert-Poincaré series of a suitable equivariant bundle on $S^d$, i.e. a manifold with a dimension $n>d$ (in our case, $n=2d$). (Incidentally, in our opinion also for the Madore-Hoppe fuzzy sphere the most natural interpretation is of a quantized phase space, because
the $\hbar\to 0$ limit of the quantum Poisson bracket endows its algebra with
a nontrivial Poisson structure.)

We understand  $\Hi_\Lambda,{\cal C}_\Lambda$ as fuzzy  ``quantized" $S^d$ in the following weaker sense. $\Hi_\Lambda,{\cal C}_\Lambda$  are the quantizations of  ${\cal L}^2(S^d),\,C(S^d)$, because, by (\ref{defhatT}b), the whole
$\Hi_\Lambda$ is obtained applying to the ground state $\bpsi_0$ 
%(or any other $\bpsi\in\Hi_\Lambda$)  
the polynomials in the $\overline{x}^i$ alone (or the subspace $\C_\Lambda$), or equivalently [by %$L_{ij}=i[X^j,X^i]$ and 
(\ref{Opcorr'})] the polynomials in the $X^i=L_{\bD i}$ alone,  in the same way as ${\cal L}^2(S^d)$ is obtained (modulo completion) by applying $C(S^d)$ or $Pol_D$, i.e. the polynomials in the $t^i=x^i/r$, to the ground state
(the constant function on $S^d$). 
These quantizations are $O(D)$-equivariant because $\Hi_\Lambda$ (resp. ${\cal C}_\Lambda$)  carries the same reducible representation of $O(D)$  as the space  $Pol_D^\Lambda$ (resp. $Pol_D^{2\Lambda}$)  of polynomials of degree $\Lambda$  (resp.  $2\Lambda$) in the $t^i=x^i/r$.  Identifying $\Hi_\Lambda,{\cal C}_\Lambda$ with $Pol_D^\Lambda,Pol_D^{2\Lambda}$ as $O(D)$-modules, as $\Lambda\to\infty$ the latter
become dense in ${\cal L}^2(S^d)$, $C(S^d)$, and their decompositions into irreps
of  $O(D)$ become that (\ref{directsum}) of both ${\cal L}^2(S^d)$,  $C(S^d)$.
This is not the case for the other fuzzy spheres. 

\smallskip
We expect that  space uncertainties and optimally localized/coherent states for $d=1,2$ \cite{FioPis20}    generalize  to $d>2$. It is also worth investigating about: distances between optimally  localized states
(as e.g. in \cite{DanLizMar14}); extending our construction to particles with spin; QFT on $S^d_\Lambda$; their application to problems in quantum gravity, or condensed matter physics; etc.  
Finally, we mention that by using Drinfel'd twists one can  construct \cite{FioreWeber,FioFraWebquadrics} a different kind of noncommutative submanifolds of noncommutative $\RR^D$,  equivariant with respect to a `quantum group' (twisted Hopf algebra).

\end{document}